# Fabrication of graphene nanodisk arrays using nanosphere lithography

C. X. Cong,[1] T. Yu,[1a)] Z. H. Ni,[1] L. Liu,[1] Z. X. Shen[1] and W. Huang[2]

[1]*Division of Physics and Applied Physics, School of Physical and Mathematical Sciences, Nanyang Technological University, Singapore 637371, Singapore*

[2]*Jiangsu Key Lab for Organic Electronics & Information Displays, Nanjing University of Posts and Telecommunications, 9 Wenyuan Road, Nanjing, China 210046*

Ordered graphene nanodisk arrays have been successfully fabricated by combining nanosphere lithography (NSL) and reactive ion etching (RIE) processes. The dimension of graphene nanodisks can be effectively tuned by varying the size of polystyrene spheres, which function as masks during RIE. Low voltage scanning electron microscopy shows that the graphene sheet could be readily patterned into periodic disk-like nanostructures by oxygen RIE. Raman mapping and spectroscopy further visualize such nanodisk arrays and reveal the nature of disks are crystalline single layer graphene. This work demonstrates an efficient and manageable way to pattern graphene. Considering the periodicity, nanometer dimension and large edge to body ratio, the graphene nanodisk arrays, such two dimensional assembly of carbon atoms offer intrisic advantages in various electronic and spintronic fabrications.

---

a) Author to whom correspondence should be addressed; E mail: yuting@ntu.edu.sg

**Introduction**

Graphene has attracted significant attention because of its unique electronic properties, such as the charge carriers mimicking massless Dirac fermions, electron-hole symmetry near the charge neutrality point, and weak spin-orbit coupling.[1,2] Graphene-based nanostructures are considered as promising candidate for an alternative to silicon based mesostructures in future electronic nanodevices. It was predicted that graphene nanoribbons with certain edge chirality would open the band gap [3,4] and show distinguish magnetic,[5] optical [6] and superconductive [7] properties. It was also reported that a finite nanostructured graphene with special edges can exhibit giant spin moments.[8] Although, graphene nanoribbons have been investigated intensively in both experimental and theoretical aspect,[9,10] graphene nanodisk, another class of graphene derivatives, which has closed edges, is still in its infancy, especially from an experimental point of view. The existence of closed edge structure may endows graphene nanodisk arrays with higher edge density than that of normal graphene sheets, leading to higher edge reactivity property, which may offer key advantages in realizing various electron applications via edge chemical functionalization, such as doping.

So far, two methods: soft-landing mass spectrometry[11] and electron-beam lithography[12] have been commonly used for patterning graphene nanostructures, such as nanoribbons, quantum dots and nanorings. However, both of them involve either rather high-cost and low-throughput lithographic patterning or sophisticated instruments, hindering their large-scale fabrication and practical applications. Till now, there is no report about patterned graphene nanodisk arrays, which would be fundamental and technological important in electronics, spintronics and photonics given their large edge to body ratio, periodicity and confinement. Here, we report a effective approach for patterning graphene sheet to ordered graphene nanodisk arrays by combining NSL with $O_2$ RIE. Raman spectroscopy/imaging and scanning electron microscopy (SEM) were employed to visualize the graphene nanodisks and reveal the nature of disks are crystalline single layer graphene. In the Raman spectra of graphene nanodisks, strong defect-related Raman bands have been found. These process-induced structural defects which were thought to be formed during the RIE process may give rise to localized electronic states, and eventually to the magnetic state in graphene based on the defect-mediated mechanism.[13]

**Experimental Methods**

Experimentally, graphene sheets were transferred from highly ordered pyrolytic graphite (HOPG, SPI SUPPLIES) by mechanical cleavage.[14] The substrate is Si wafer coated by 300 nm thermally-grown $SiO_2$. Polystyrene (PS) spheres of various diameters (Polysciences, Inc) were used as mask. The RIE process was performed by using March PX-250 plasma etching system with power of 70 W and base pressure of 70 mTorr. Pure oxygen gas was used as plasma source. The pristine graphene sheets and graphene nanodisks were characterized by optical microscopy, low voltage scanning electron microscopy (SEM, JOEL JSM-6700F) and Raman mapping/spectroscopy (WITEC CRM200 confocal system, $\lambda_{laser}$ = 532 nm). To avoid heating effect, the laser power at sample surface was fixed at 0.5 mW. The details about Raman imaging has been explained in Ref. 19.

**Results and Discussion**

Figure 1 shows the schematic of the graphene nanodisk arrays fabrication process. Firstly, the graphene sheets were transferred to $SiO_2$/Si substrate (Fig. 1(a)). Optical microscopy was used to locate the interested thin graphene sheets where the single-layer graphene sheet was further confirmed by Raman spectra. Secondly, a monolayer of highly ordered PS spheres was self-assembled on water surface using a technique reported by J. Rybczynski et al.[15] Such monolayer was then lifted off from the water surface using previously mentioned $SiO_2$/Si substrate with graphene (Fig. 1 (b)). Then, $O_2$ RIE was carried out to morph the closely packed PS nanosphere monolayer into arrays of separated nanospheres, and to etch a portion of the graphene sheets that was not protected by the nanospheres (Fig. 1 (c)). Finally, the PS spheres were removed by sonication in chloroform for a few seconds, resulting in periodic graphene nanodisk arrays (Fig. 1 (d)). To clean the graphene nanodisks, a vacuum (x$10^{-3}$ Torr) post-annealing process was conducted at 500 °C for 30 min by a Linkam thermal stage.

Optical and SEM images are shown in Fig. 2 to demonstrate the result of each step described above. Fig. 2(a) shows the graphene sample on $SiO_2$/Si substrate. The quadrilateral area marked by dashed line is single-layer graphene sheet. Figs. 2(b) and 2(c) show the optical and high magnification SEM images of single-layer graphene covered by monolayer of PS spheres before and after $O_2$ RIE. It can be seen from the SEM images (Figs. 2(b) and 2(c) insets) that the size of PS spheres was reduced from 465 to about 450 nm after $O_2$ RIE. Therefore, each PS sphere can be distinguished more clearly than those before $O_2$ RIE, as shown in the optical images. The graphene sheet below the gaps of spheres was therefore directed exposed to the $O_2$ plasma and selectively etched away. Highly ordered arrays of graphene nanodisks are obtained after removing the PS spheres, whose optical image is basically the same as that of the as-transferred graphene (not shown here). To prove the formation and reveal the morphologies of graphene nanodisk arrays, SEM was performed. As $SiO_2$ substrate is insulative, a low acceleration voltage of 1 kV was set during SEM measurement to minimize the charging effect and have a good contrast. As shown in Fig. 2(d), the graphene nanodisk arrays were readily formed and the average diameter is about 448 nm, which agrees well to the size of spheres after the $O_2$ RIE. Therefore, we concluded that by combining NSL with $O_2$ RIE the graphene nanodisk arrays can be easily obtained. Furthermore, the diameter of the graphene nanodisk and the distance between two adjacent graphene nanodisks can be well tuned because the diameter of PS spheres and the distance between two adjacent PS spheres can be controlled by adjusting RIE conditions such as etching time, power, pressure, as well as initial size of the PS spheres.[16]

Figure 3 shows the relationship between graphene nanodisk size and initial PS sphere size. The insets of Fig. 3 are SEM images of different size graphene nanodisk arrays fabricated by using different size of PS spheres. The graphene nanodisk sizes are 448 ± 17 nm, 316 ± 12 nm, 186 ± 10 nm, which are fabricated by using initial size of 465 ± 11 nm, 356 ± 14 nm, 202 ± 10 nm PS sphere as masks, respectively, under the same oxygen RIE conditions. It can be seen that the diameter of the graphene nanodisk can be well tuned by changing the initial PS sphere size. Therefore, the size of graphene nanodisks can be scaled down to tens of nanometers by using tens of nanmeters PS spheres as mask. The fabrication and properties of the smaller graphene nanodisk arrays will be studied in our future work.

To understand the properties and structure of the obtained samples, Raman spectroscopy studies are also carried out. Raman spectroscopy is one of the most powerful techniques capable of probing many properties of graphene such as identifying number of layers, sensing structural defects, revealing doping impurities, and detecting strain. [17-22] The Raman spectra of the as-transferred graphene and the graphene nanodisk arrays with different sizes are shown in Fig. 4(A). The sharp 2D band of the pristine single-layer graphene [18] can be clearly observed in the Raman spectrum. The Raman spectra of graphene nanodisk arrays with different sizes are similar. A clear difference between the spectra of the graphene nanodisks and the original graphene is that an extra Raman band, locating at 1340 cm$^{-1}$ presents in the spectra of the graphene nanodisks. This peak corresponds to the so-called disorder induced D band, which is activated by a double resonance effect by defects, such as vacancies, or grain boundaries, and so on. [23] Another difference is that the G band of graphene nanodisk arrays becomes broader than that of the as-transferred graphene. It can be attributed to that the G band was composed of two peaks, which are locating at 1594 cm$^{-1}$ and 1620 cm$^{-1}$, respectively. The shoulder peak locating at 1620 cm$^{-1}$ is also related to defects, commonly called D′ band. The observation of D and D′ bands indicated that many defects were introduced into the graphene nanodisk arrays by $O_2$ RIE. The G band located at 1594 $cm^{-1}$ of the graphene nanodisks shifted 14 $cm^{-1}$ towards high frequency with respect to that of the as-transferred graphene (1580 $cm^{-1}$), meanwhile the 2D band exhibited 15 $cm^{-1}$ blue-shift as well. These blue-shifts might be explained by considering the charge doping by oxygen atoms or ions binding with the carbon dangling bonds during oxygen RIE process. It was reported that besides the G band blueshift, a bandwidth narrowing was also observed in the case of charge doping.[20] However, in our results, the bandwidths of both G and 2D are broader than those of the as-transferred graphene, which might be due to the defect-induced disordering in graphene nanodisks. Typical Raman images of the graphene nanodisk arrays fabricated by using monolayer of 465 nm PS spheres as masks are shown in Fig. 4(B1-B3). For the graphene nanodisk arrays, it can be seen that the Raman images (plotted by extracting the integrated intensity of D band, G band, and 2D band) present the good periodic structures which agree very well with the optical and SEM images.

The high edge density existing in the graphene nanodisk arrays may lead to high edge reactivity property as well as high edge localized spin-polarized electronic states, which may offer key advantages in various spintronic applications and edge chemical functionalization. Moreover, it has been calculated that the vacancy defects such as single-atom void,[24] two-atom voids with the same spin brought together, and sufficiently large voids with sublattice imbalance of zero [25] can induce local magnetism in graphene-based materials. As normally-presented defects in carbon clusters, the carbon vacancy complex mentioned above would of course present in the graphene nanodisk arrays and leads to the magnetic state as well. Therefore, our graphene nanodisk system would be an ideal structure to examine the correctness of such novel physics idea. Accordingly, further studies of the graphene nanodisks on their defect-induced and edge-induced magnetism deserve to carry out at atomic level by scanning tunneling microscopy (STM) and transmission electron microscopy (TEM).

**Conclusion**

In conclusion, graphene nanodisk arrays were fabricated by combining NSL with $O_2$ reactive ion etching for the first time. The size of the graphene nanodisk can be easily tuned by changing the diameters of PS spheres. Low voltage SEM and Raman mapping were employed to visualize the graphene nanodisk arrays. In the Raman spectra of graphene nanodisks strong defect bands were observed. The defects introduced in graphene nanodisks may generate the magnetic state in graphene, which would be important in spintronics application. Moreover, graphene nanodisk arrays as a close-edged nanostrusture having the high edge reactivity property can offer key advantages in edge chemical functionalization.

**Figure captions**

**Figure 1**. Schematic of graphene nanodisk arrays fabrication process. (a) The graphene sheets were transferred to SiO$_2$/Si substrate. (b) The PS speres were arranged on the SiO$_2$/Si substrate with graphene. (c) O$_2$ RIE was carried out to etch a portion of the graphene sheets that was not protected by the PS spheres. (d) Removal of PS spheres.

**Figure 2**. Optical and SEM images of (a) graphene on SiO$_2$/Si substrate; (b) graphene covered by monolayer of PS spheres of 465 nm in diameter; (c) graphene nanodisk with PS monolayer after oxygen RIE; (d) graphene nanodisk after remove the PS monolayer. The insets of (b) and (c) are SEM images of graphene covered by monolayer of PS spheres of 465 nm in diameter before and after oxygen RIE, respectively. (The scale bar of the inset SEM images is 500 nm.)

**Figure 3.** Plot of graphene nanodisk size as a function of PS sphere size and inset SEM images of different size of graphene nanodisks fabricated by different sizes of PS spheres: 202 nm(a), 306 nm(b), and 465 nm(c). (The scale bar of the inset SEM images is 500 nm.)

**Figure 4**. (A) Raman spectra of curve (a) as-transferred graphene on SiO$_2$/Si substrate; (b) 448 nm graphene nanodisk arrays; (c) 316 nm graphene nanodisk arrays; (d) 186 nm graphene nanodisk arrays. The fitted G and D' peaks presenting in the Raman spectra of the graphene nanodisks were also illustrated. And Raman images of 448 nm

graphene nanodisks plotted by extracting the integrated intensity of D band (B1), G band (B2), and 2D band (B3).

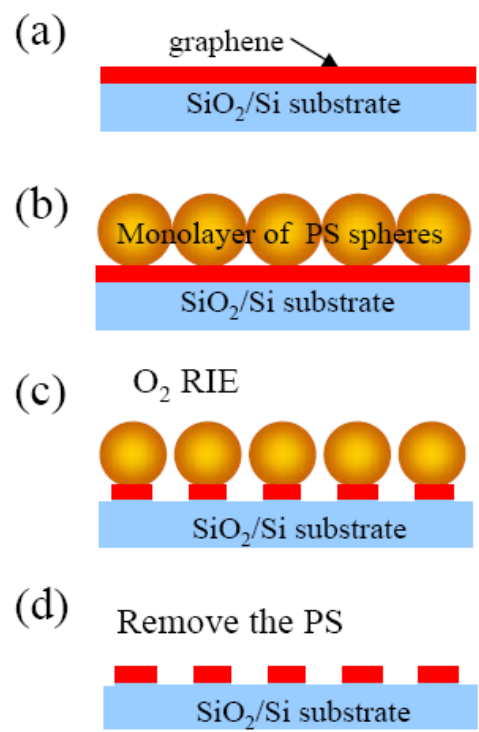

Fig. 1

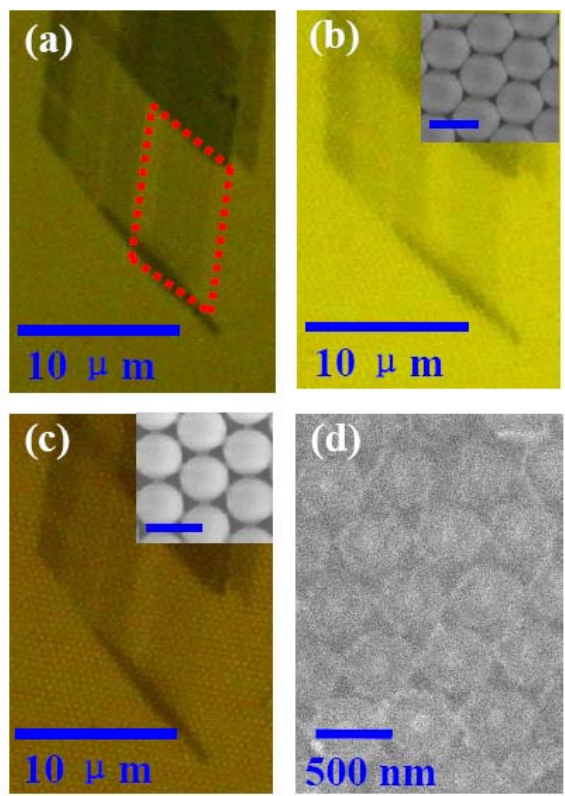

Fig. 2

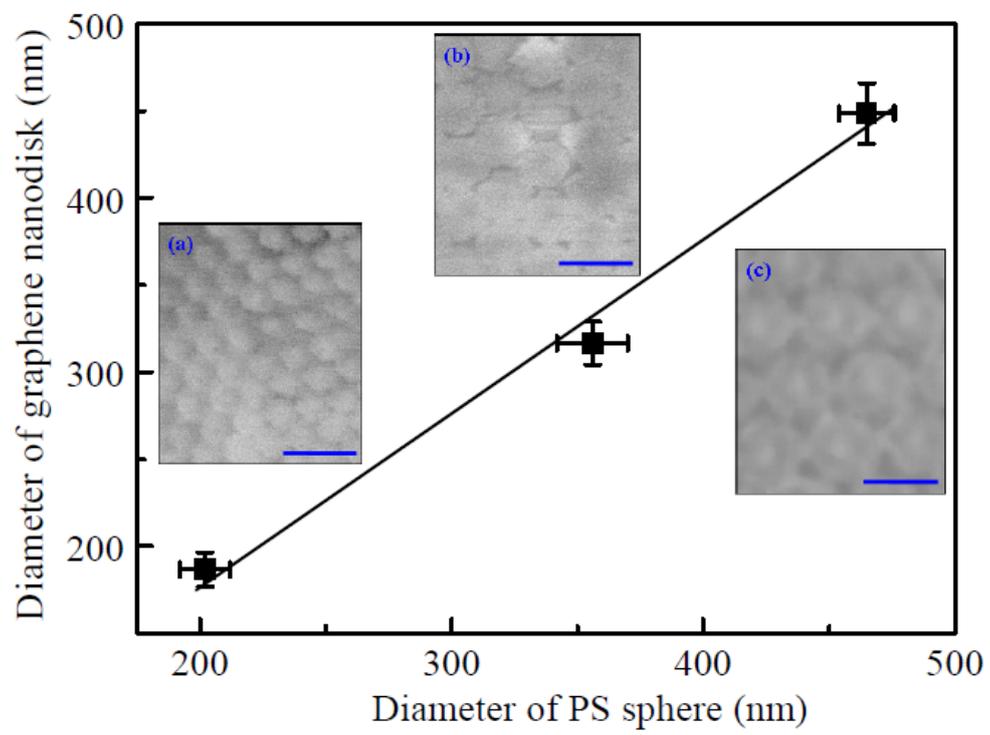

Fig. 3

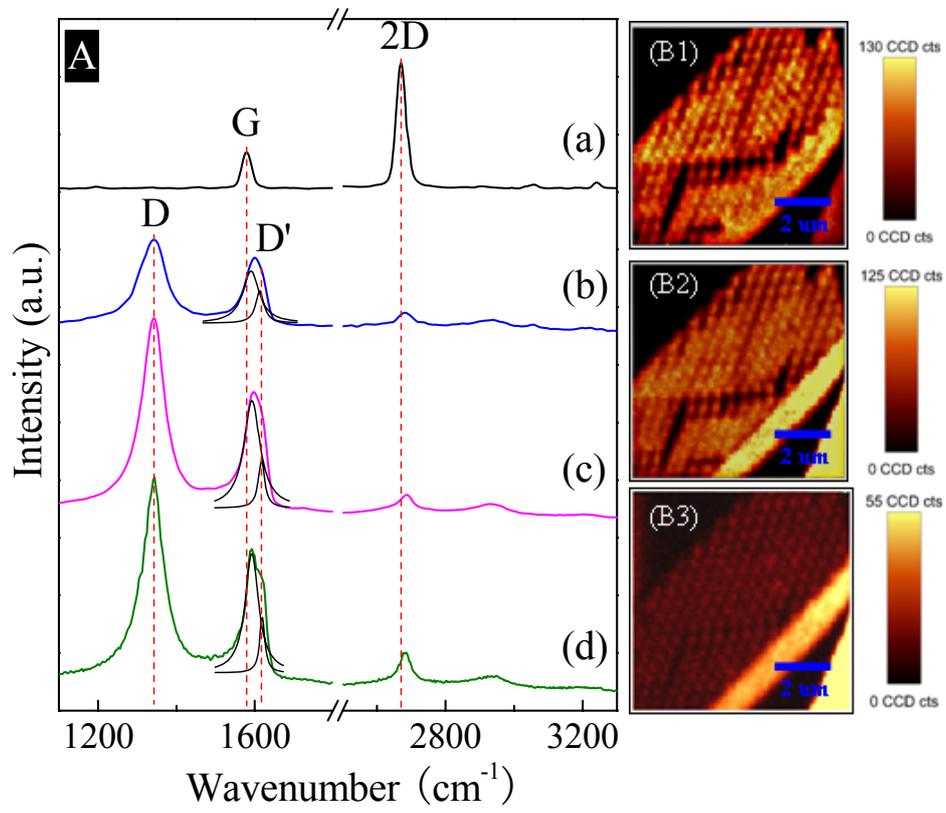

Fig. 4